**Title**: Variability analysis in memristors based on electrodeposited prussian blue

**Authors:** L. B. Avila[1], A. Cantudo[2], M.A. Villena[2], D. Maldonado[2,3], F. Abreu Araujo[1], C. K. Müller[4], J. B. Roldán[2,5]

**Address:**

[1]Institute of Condensed Matter and Nanosciences (ICMN), Université catholique de Louvain (UCLouvain), Louvain-la-Neuve B-1348, Belgium.

[2]Departamento de Electrónica y Tecnología de Computadores, Universidad de Granada, Facultad de Ciencias, Avd. Fuentenueva s/n, 18071 Granada, Spain.

[3]Departamento de Informática y Estadística, Esc. Tec. Sup. de Ingeniería Informática, Universidad Rey Juan Carlos, Móstoles, 28933, Madrid, Spain.

[4]Faculty of Physical Engineering/Computer Sciences, University of Applied Sciences Zwickau, 08056 Zwickau, Germany

[5]Institute "Carlos I" for Theoretical and Computational Physics, University of Granada, 18071, Granada, Spain

Corresponding author email: jroldan@ugr.es

**Abstract:**

This work presents a comprehensive analysis of the variability and reliability of the resistive switching (RS) behavior in Prussian Blue (a mixed-valence iron(III/II) hexacyanoferrate compound) thin films, used as the active layer. These films are fabricated through a simple and scalable electrochemical process, and exhibit robust bipolar resistive switching, making them suitable both for neuromorphic computing applications and hardware cryptography. A detailed statistical evaluation was conducted over 100 consecutive switching cycles using multiple parameter extraction techniques to assess cycle-to-cycle (C2C) variability in key RS parameters, including set/reset voltages and corresponding currents. One and two-dimensional coefficients of variation (1DCV and 2DCV) were calculated to quantify variability and identify application potential. Results demonstrate moderate variability compatible with neuromorphic computing and cryptographic functionalities, including physical unclonable functions and true random number generation. These findings position Prussian Blue-based memristors as promising candidates for low-cost, stable, and multifunctional memory.

# 1.-Introduction

Memristors, a new kind of electron devices [Chua1971, Chua1976] that shows hysteresis in their current versus voltage curves linked to its materials properties, are drawing attention from the electronics industry due to the great potential of their outstanding applications [Lanza2022, Lanza2025]. The most straightforward use of these devices is found in the non-volatile memory realm [Wong2012, Chiu2019, Spiga2020]. The resistive switching (RS) features of memristors allow multilevel operation [Reuben2019, Perez-Bosch2021, Poblador2018], as is the case in conventional Flash technology. In this respect, different foundries and IC companies, such as TSMC and INTEL, have shown the correct operation of non-volatile memory chips in the context of the 22nm technology [Chou2020, Jain2019]. In the field of cryptography, the inherent stochastic behavior of these devices allows the fabrication of hardware solutions for the implementation of entropy sources for true random number generators [Wen2021], or physical unclonable functions [Nili2018]. In addition, the memristive device internal dynamics can mimic the operation of biological synapses, accounting for the ionic transport that triggers RS [Tang2019, Roldan2022, Mishchenko2022, Maestro-Izquierdo2019, Mikhaylov2020]. Consequently, these devices play a key role in neuromorphic engineering, a subject introduced in the late eighties by Carver Mead [Mead1989]. Memristors are essential in the implementation of crossbar arrays that can optimize vector-matrix multiplication, a crucial operation for the acceleration of training and inference processes in hardware neural networks [Ambrogio2018, Milo2016, Prezioso2015, Sebastian2020, Aguirre2024, Hui2021, Danilin2015, Perez-Bosch2021, Yu2021, Zhu2023, Wang2020].

Despite their game-changing applications, some issues need to be addressed to improve memristor technologies, e.g. the lack of accurate modeling and simulation tools to aid in the device and circuit design [Lanza2019, Roldan2021]. Variability, both in the cycle-to-cycle [Roldan2022b, Roldan2023, Grossi2016, Maldonado2022] and device-to-device [Perez2019] flavors, also needs improvement. In particular, in what is connected to the statistical analysis of data and in compact modeling to build reliable circuit simulators and other EDA tools for integrated circuit design. In order to reduce variability and advance in other facets such as endurance, retention, etc., different materials have been employed to build the device stack. In this context, new electrodes and mostly, dielectrics, are being studied, such as transition-metal oxides [Lanza2022, Spiga2020, Poblador2020, Pan2014], two-dimensional (2D) layered materials, boron-based compounds [Knot2022], MXenes [Knot2021], perovskites and nanotubes [Nirmal2022, Bisquert2024], quantum dots and polymers [Sangman2020, Avila2024b], etc.

Resistive Random-Access Memories (RRAM) are counted among memristive devices [Chua2011]. Their simple metal-oxide-metal structure allows them to be fabricated in the BEOL of CMOS integrated circuits [Zhu2023]. In the context highlighted above, we present here a variability study of RRAMs where a new material is considered as the dielectric, the Prussian Blue (PB) [Avila2020, Faita2021, Avila2022, Avila2024].

Prussian Blue is a mixed-valence iron(III/II) hexacyanoferrate compound, renowned for its distinctive cubic face-centered structure. In its crystal lattice, $Fe^{3+}$ ions are coordinated by nitrogen atoms, while $Fe^{2+}$ ions are coordinated by carbon atoms, forming the repeating bonding chain $Fe^{3+}-N\equiv C-Fe^{2+}-C\equiv N-Fe^{3+}$. This well-ordered structure not only causes the material's characteristic deep blue color but also endows it with unique physical and chemical properties [Buser1977]. Substitution of iron ions with other transition metal ions (e.g. cobalt or manganese) gives rise to a broader class of materials known as Prussian Blue analogs (PBAs) [Brown1969]. Electrodeposition is a highly versatile technique for depositing PB onto conductive substrates from an aqueous electrolyte solution. This method offers exceptional flexibility, as it is compatible with various metals and adaptable to different device sizes. Such advantages make it particularly suitable for wafer-scale device integration. In this work, Prussian Blue thin films were synthesized via electrodeposition, and their morphological and electrical properties were systematically investigated to evaluate their potential for device applications. An in-depth variability study is performed from the statistical viewpoint.

The device fabrication details, and measurement setup are described in section 2, the variability statistical study is given in section 3, and the conclusions are drawn in section 4.

## 2.-Device fabrication and measurement setup

The electrochemical deposition of prussian blue (PB) films was carried out in potentiostatic mode using an electrochemical workstation (Ivium CompactStat, Eindhoven, Netherlands) at room temperature. The electrochemical system used for the fabrication of PB is presented in Figure 1a, where the working electrode (WE) consists of Au/Cr/Si substrates. At the same time, a Pt foil served as the counter electrode (CE) and a saturated calomel electrode (SCE), as the reference electrode (RE), all connected to the Potentiostat. The working electrode was fabricated by evaporating 50 nm of Au onto a 5 nm Cr layer on (100) Si substrates (1 cm × 1 cm) at a base pressure of $10^{-5}$ Pa. PB deposition occurred within a 0.5 cm² circular area, defined by an adhesive tape mask.

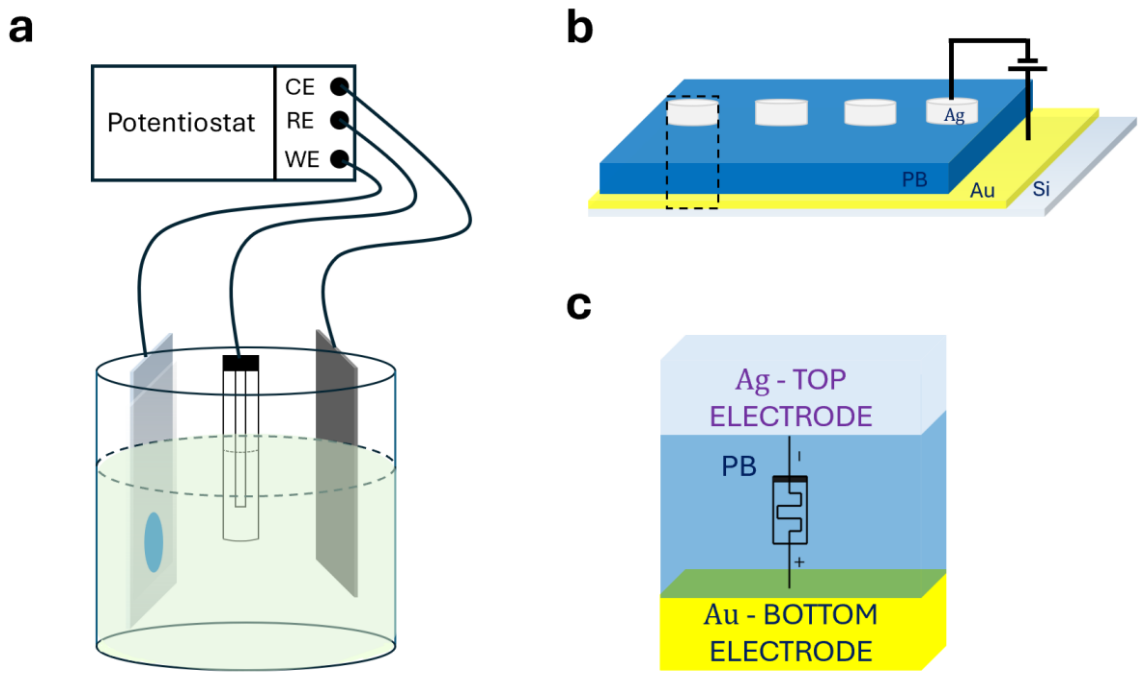

**Figure 1. a** Schematic of the experimental setup for the electrochemical deposition of PB thin films, showing the Counter Electrode (CE, Pt), Reference Electrode (RE), Working Electrode (WE), and the electrolyte (depicted in green). **b** Schematic representation of the sample's layer sequence and electrode configuration used for electrical measurements. **c** Close-up view of a single PB vertical memristor.

The electrolyte solution for electrochemical synthesis consisted of 1.0 M KCl, 5.0 mM HCl, 0.5 mM $FeCl_3$, and 0.5 mM $K_3Fe(CN)_6$, all dissolved in 100 mL of deionized water, with the pH adjusted to 2. All chemicals (Sigma Aldrich, Darmstadt, Germany) had a purity of >99%. PB layers were deposited by applying a constant potential between the WE and CE of 0.3 V at 25°C, with the electrodeposited charge limited to 30 mC, with a thickness of 500 nm. The sample morphology was characterized using field emission scanning electron microscopy (FEG-SEM; TESCAN CLARA, Brno, Czech Republic). As shown in Figure 2, the PB film electrodeposited exhibits a well-defined cubic crystal morphology, with a highly ordered structure matching PB's intrinsic face-centered cubic (fcc) structure. The film displays a compact and homogeneous grain distribution across the substrate (Inset photo). Small clusters composed of aggregated cubes were also observed on the surface (Figure 2).

The study of these devices based on Prussian blue is at the prototype level currently, further improvements would be needed to face the issues related to wafer level fabrication.

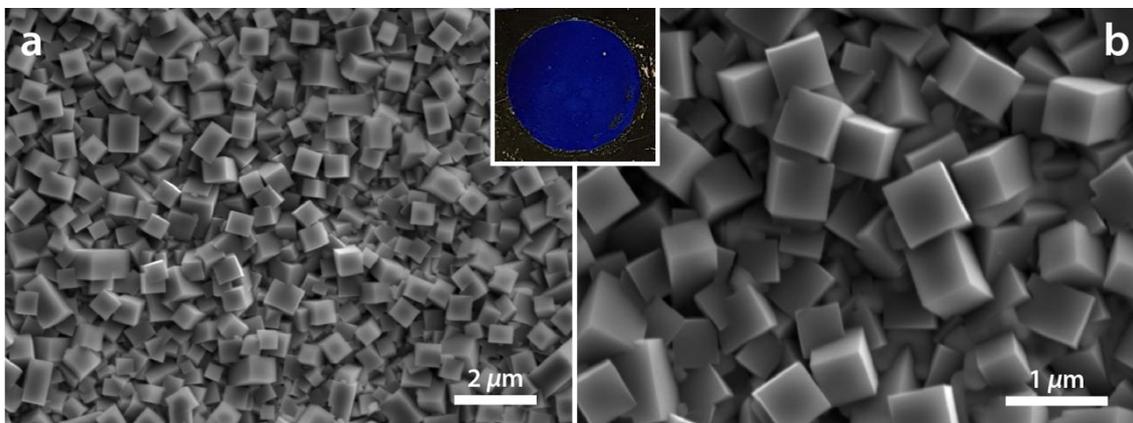

**Figure 2.** SEM images of the PB layer electrodeposited at 0.3 V obtained at 10 keV. **a** surface overview, photograph image of the film (inset), **b** a more detailed view of the film surface.

To investigate the resistive switching effect, current-voltage (I-V) characterization was performed using a Keithley 2400 programmable electrometer. The measurements were conducted with a two-point probe configuration. The device structure consisted of a PB film sandwiched between a top silver (Ag) electrode and a bottom gold (Au) electrode, forming a metal-insulator-metal (MIM) structure essential for resistive switching characterization. The bottom Au electrode served as the fixed contact, providing a stable and uniform interface for the PB film, while the top Ag electrode was prepared by drop casting directly applied onto the PB surface. Figure 1b illustrates the complete structure, while Figure 1c presents the structure of a single memristor.

Electrical characterization, specifically current-voltage (I-V) measurements, was performed on the devices using the same electrochemical workstation used for electrodeposition. A DC voltage sweep ranging from −1 V to +1 V was applied at a scan rate of 5 mV/s during the measurement. Figure 3a displays the current as a function of applied voltage under the ramp voltage stress regime, revealing a characteristic butterfly-shaped curve. The solid line represents the first cycle, showing the forming free characteristics of these devices. The dash line corresponds to the 100$^{th}$ cycle, highlighting the device stability and repeatability over multiple cycles. In the positive polarity, the transition from the low-resistance state (LRS) to the high-resistance state (HRS) is observed, indicating the *Reset* process. The device remains in the HRS until the polarity is reversed. Upon applying negative bias, the transition from HRS to LRS occurs, corresponding to the *Set* process.

As shown in Figure 3b, the current transient and applied voltage are plotted as a function of time, a complete switching cycle takes 800 seconds. In this study, we employ slower scan rates to minimize device stress, ensuring more reliable and repeatable results, an essential factor for assessing long-term stability and endurance. Additionally, a reduced scan rate helps suppress capacitive and transient effects, making the resistive component of the current more prominent. By allowing sufficient time for the necessary physical or chemical changes to occur, the slower scan rate enables a more accurate observation of the device's intrinsic switching behavior, leading to a clearer representation of its switching kinetics.

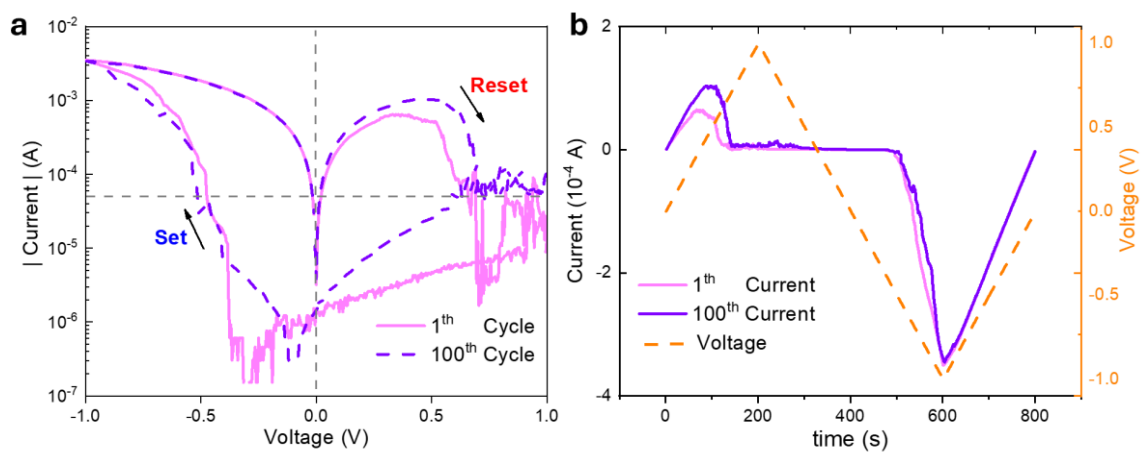

**Figure 3. a** Representative resistive switching I-V curve illustrates the transition between HRS and LRS. The magenta curve represents the 1st measurement, while the violet dashed curve corresponds to the 100th measurement. **b** The graph shows the current and voltage transients over time.

## 3.-Variability statistical study

As shown in Figure 3a, these devices exhibit two well-defined states (HRS and LRS), making them good candidates for non-volatile memories or neuromorphic applications. However, the C2C variability requirements for each of these applications are different. Therefore, a thorough analysis of their variability is required to accurately identify this technology potential.

Figure 4 shows the I-V curves in a RS series for the devices under study. The set voltages ($V_{set}$) and reset voltages ($V_{reset}$) are marked with red crosses in the curves, 100 complete I-V curves have been analyzed. The corresponding current values, $I_{set}$ and $I_{reset}$, are found in the experimental I-V curves and correspond to $V_{set}$ and $V_{reset}$.

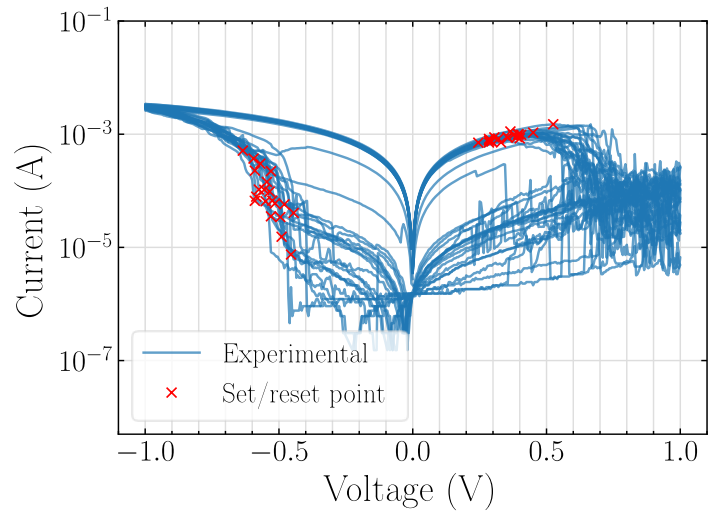

**Figure 4.** Current versus voltage for the devices under study. The set and reset voltages are shown in red crosses in the curves. The ramped voltage employed in the measurements had a slope of 5 mV/s.

For the RS parameter extraction, we employed three different numerical procedures, see Ref. [Maldonado2022b, Villena2014]. $V_{set}$ was obtained by means of (MS1 technique in Maldonado2022b) the determination of the maximum current increase (35%) between two consecutive current points, the MS2 technique extracts the set I-V curve elbow; finally, the MS3 technique is based on the determination of the maximum current derivative along the curve, see Figure 5a. $V_{reset}$ was obtained (MR1) through the first current decrease along the I-V curve, the MR2 technique is connected to the minimum (or null) current derivative along the current curve, and the MR3 technique extracts the maximum current decrease (80%) in two consecutive current points, see Figure 5b.

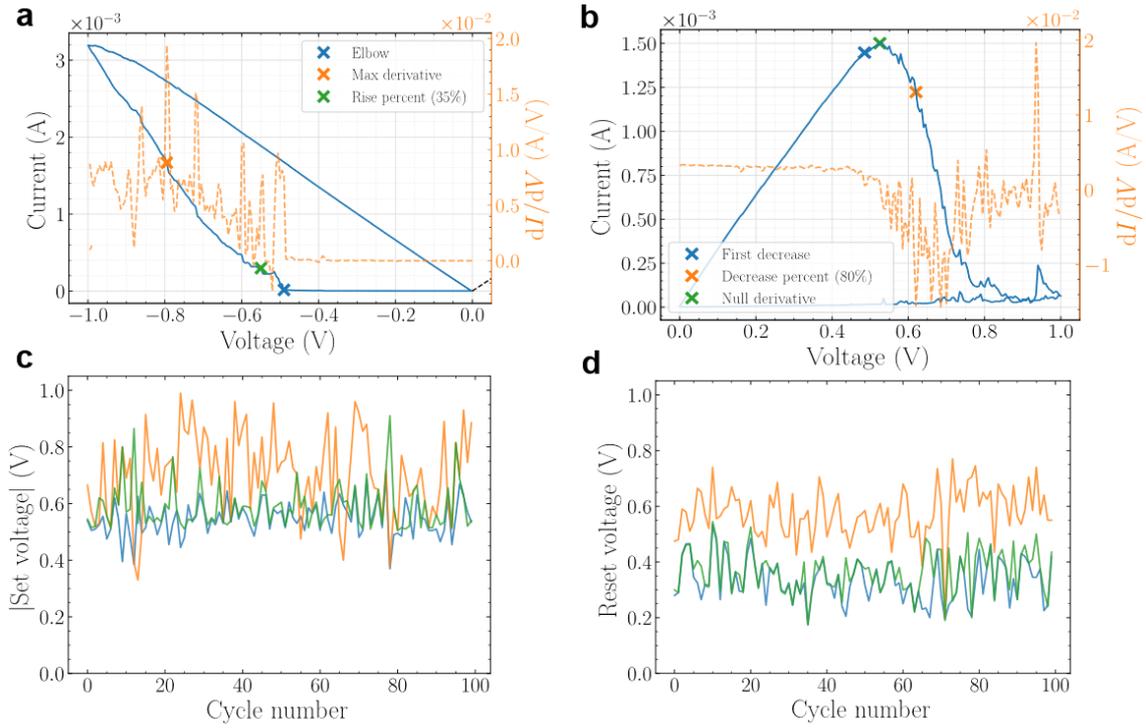

**Figure 5. a** Set current versus voltage curve where the three set voltage extraction procedures employed are shown [Maldonado2022b]. Determination of the maximum current increase (35%) between two consecutive current points (green points, MS1), maximum separation of the experimental I-V curve with a straight line that joins the first and final points of the I-V curve (blue points, MS2, Elbow method) and maximum current derivative in the current curve (orange points, MS3). **b** Reset current versus voltage curve showing the three reset voltage extraction procedures used [Maldonado2022b]. Calculation of the first current decrease along the I-V curve (blue points, MR1), minimum (or null) current derivative along the curve (green points, MR2), and maximum decrease (20%) in two consecutive current points (orange points, MR3). **c (d)** Set (reset, absolute value) voltage versus the number of cycles for the prussian blue based devices.

Making use of these data, an in-depth statistical study on cycle-to-cycle (C2C) variability is performed by using the coefficient of variation (CV) in several dimensions [Acal2024]. The one-dimensional version of this coefficient (1DCV, defined as $\sigma/\mu$, where $\sigma$ stands for the standard deviation and $\mu$ is the mean) and the two-dimensional CV (2DCV) [Acal2023, Acal2024] are shown in Table 1, calculated considering the pairs of values ($V_{reset}$, $I_{reset}$) or ($V_{set}$, $I_{set}$). The 1DCV has been used in the past for the analysis of the C2C variability. For 1DCV($V_{set}$) < 2%, the devices are recommended for information storage [Chiu2019];

although higher values could be employed in the neuromorphic computation context [Sebastian2020]. For 1DCV($V_{set}$) > 20%, the devices could be employed for data encryption, as an entropy source of true random number generators [Wen2021], or physical unclonable functions [Nili2018].

The 2DCV is higher for the set voltage (Table 1c) with respect to the reset voltage (Table 1d). Nevertheless, the 1DCV for the set voltage (Table 1a) is lower, in general, than for the reset voltage (Table 1b). These statistical effects (different trends for the 1DCVs and 2DCVs) have been observed previously [Acal2024]; in this respect, a 2D dataset produces a much more accurate analysis.

**a** 1D CV, Set

| Method | CV |
|---|---|
| Elbow | 0.113 |
| Max derivative | 0.207 |
| Rise percent (35%) | 0.136 |

**b** 1D CV, Reset

| Method | CV |
|---|---|
| First decrease | 0.228 |
| Decrease percent (80%) | 0.152 |
| Null derivative | 0.213 |

**c** 2D CV, ($V_{\text{Set}}$, $I_{\text{Set}}$)

| Method | CV |
|---|---|
| Elbow | 0.070 |
| Max derivative | 0.160 |
| Rise percent (35%) | 0.093 |

**d** 2D CV, ($V_{\text{Reset}}$, $I_{\text{Reset}}$)

| Method | CV |
|---|---|
| First decrease | 0.078 |
| Decrease percent (80%) | 0.089 |
| Null derivative | 0.080 |

**e** 2D CV, ($V_{\text{Set}}$, $V_{\text{Reset}}$)

| Method | | CV |
|---|---|---|
| Elbow | First decrease | 0.100 |
| Elbow | Decrease percent (80%) | 0.109 |
| Elbow | Null derivative | 0.101 |
| Max derivative | First decrease | 0.169 |
| Max derivative | Decrease percent (80%) | 0.174 |
| Max derivative | Null derivative | 0.170 |
| Rise percent (35%) | First decrease | 0.113 |
| Rise percent (35%) | Decrease percent (80%) | 0.120 |
| Rise percent (35%) | Null derivative | 0.114 |

**Table 1. a (b)** Tables with the 1DCV for the set (reset) voltages for prussian blue devices. **c (d)** Tables with the 2DCV for the set (reset) voltages for prussian blue devices. The pairs of variables employed were ($V_{set}$, $I_{set}$) and ($V_{reset}$, $I_{reset}$). **e** Table with the 2DCV for ($V_{set}$, $V_{reset}$). The CV are obtained for the different extraction techniques employed in the study [Maldonado2022b].

According to the results shown in Table 1, our devices could be used for neuromorphic computing and cryptographic applications in the current development state [Roldan2022b]. It is important to highlight that, in the context of neuromorphic computing, variability can be beneficial to overcome problems such as a lack of precision in neural network recognition due to overfitting [Romero-Zaliz2021].

A different perspective can be obtained by studying the area included between the LRS and HRS parts of the I-V curve (see Figure 6a). The evolution of this area in the RS series is shown in Figure 6b. A nonzero area within the RS loop assures the device I-V curve hysteresis and, consequently, the memory effects [Biolek2014]. This parameter could be employed for modeling purposes [Biolek2012, Biolek2014] since it is related to the action potential, which was introduced by Leon Chua in his original work [Chua1971]. See that the area value through the RS series is maintained reasonably well (save variability effects) (Figure 6b); therefore, the non-volatile memory potential does not decline as the number of cycles increases, as required for memory applications.

A more conventional analysis to assess the appropriateness of this technology for non-volatile memory applications is linked to the evaluation of the $R_{off}/R_{on}$ ratio, measured at ±0.3 V (Figure 6c). Although a 10-factor is achieved on average, an improvement in the C2C variability is still needed because for some cycles the $R_{off}/R_{on}$ ratio falls below 1. The cumulative distribution functions are shown in Figure 6d to shed light on this issue.

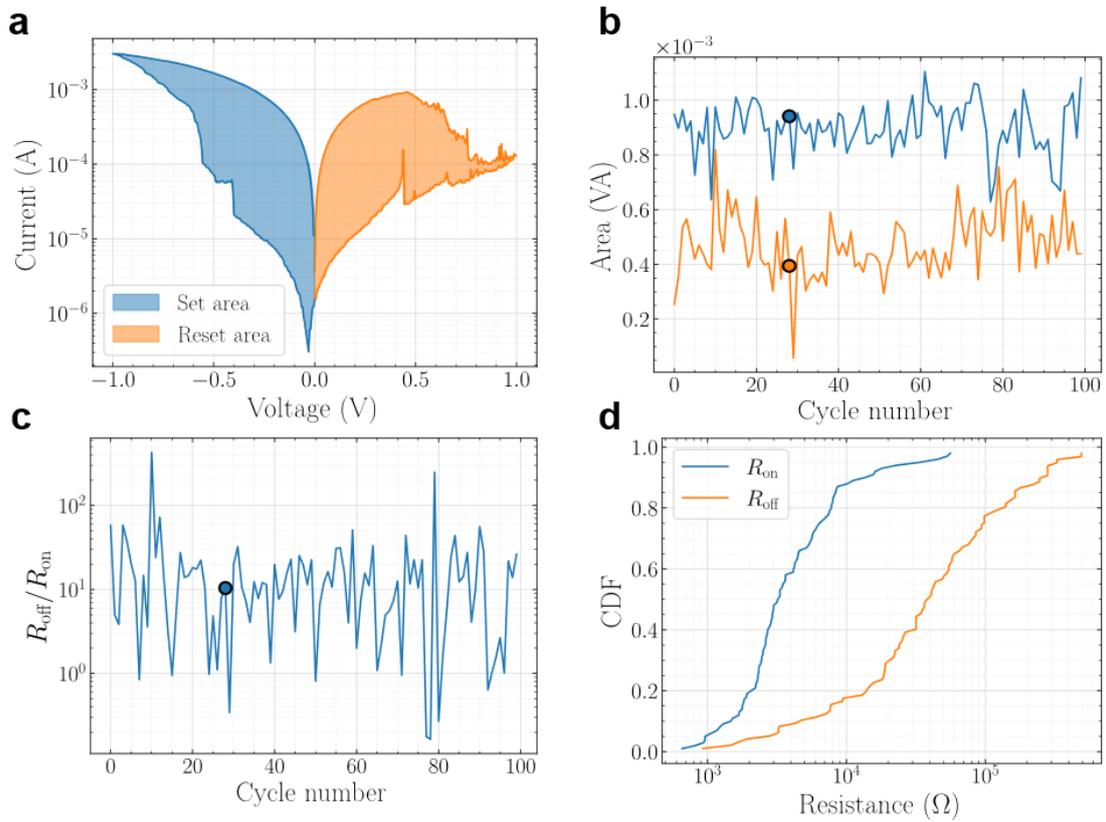

**Figure 6.** Current versus voltage curves measured under RVS. **a** The area included within the curves is shown in blue for the set process and orange for the reset process. **b** Area included in the I-V curves versus cycle number for the RS series we have measured. The dots mark the cycle corresponding to panel a **c** $R_{off}/R_{on}$ ratio versus cycle number. The dot marks the cycle corresponding to the I-V curves of panel a. **d** Cumulative distribution function for the $R_{off}$ and $R_{on}$. $R_{on}$ was extracted in the LRS for a voltage of -0.3 V, $R_{off}$ was obtained in the HRS for a voltage of 0.3 V.

## 6.-Conclusions

In this study, we have demonstrated the feasibility of using electrodeposited Prussian Blue thin films as an active material for memristive devices, highlighting their potential for resistive switching applications. Through an extensive statistical variability analysis over 100 switching cycles, we identified stable bipolar RS behavior and extracted critical performance parameters using multiple data processing techniques. The calculated coefficients of variation reveal that these devices are suitable for applications in neuromorphic computing and hardware-based cryptography. The consistent switching area and stable hysteresis loops observed over multiple cycles show the great potential of Prussian Blue devices for resistive switching operation.

## Data Availability

The datasets generated and/or analyzed during the current study are available from the corresponding author on reasonable request.

## Acknowledgements

Research supported by the project PID2022-139586NB-C44, funded by MCIN/AEI/10.13039/501100011033 and FEDER, EU; and also supported by the Ramón y Cajal grant (RYC2022-035618-I), funded by

MCIU/AEI/10.13039/501100011033, and by the FSE+. F.A.A. is a Research Fellow of the F.R.S.-FNRS. C.K.M. thanks the Deutsche Forschungsgemeinschaft for funding (No. 531524052).

**Competing Interests**

The authors declare no competing interest.

# References

[Avila2020] L. B. Avila, C. K. Müller, D. Hildebrand, F.L. Faita, B. F. Baggio, C.C. P. Cid, A.A. Pasa, "Resistive Switching in Electrodeposited Prussian Blue Layers", *Materials*, *13*, 5618, 2020. https://doi.org/10.3390/ma13245618.

[Avila2022] L. B. Avila, P.C. Serrano Arambulo, A. Dantas, E.E. Cuevas-Arizaca, D. Kumar, C.K. Müller, "Study on the Electrical Conduction Mechanism of Unipolar Resistive Switching Prussian White Thin Films", *Nanomaterials*, *12*, 2881, 2022. https://doi.org/10.3390/ nano12162881.

[Avila2024] L. B. Avila, P.A. Serrano, L.T. Quispe, A. Dantas, D.P. Costa, E.E.C. Arizaca, D.P.P. Chávez, C.D.V. Portugal, C.K. Müller, "Prussian Blue Anchored on Reduced Graphene Oxide Substrate Achieving High Voltage in Symmetric Supercapacitor", *Materials*, *17*, 3782, 2024. https://doi.org/10.3390/ ma17153782.

[Avila2024b] L. B. Avila, P. Chulkin, P.A. Serrano, J.P. Dreyer, M. Berteau-Rainville, E. Orgiu, L.D.L. França, L.M. Zimmermann, H. Bock, G.C. Faria, J. Eccher, I.H. Bechtold. Perylene-Based columnar liquid Crystal: Revealing resistive switching for nonvolatile memory devices. Journal of Molecular Liquids, 402, 124757. https://doi.org/10.1016/j.molliq.2024.124757.

[Acal2023] C. Acal, D. Maldonado, A. M. Aguilera, K. Zhu, M. Lanza, J. B. Roldán, "Holistic variability analysis in resistive switching memories using a two-dimensional variability coefficient", ACS Applied Materials & Interfaces, 15, 15, 19102–19110, 2023.

[Acal2024] C. Acal, D. Maldonado, A. Cantudo, M.B. González, F. Jiménez-Molinos, F. Campabadal, J. B. Roldán, "Variability in $HfO_2$-based memristors described with a new bidimensional statistical technique", Nanoscale, 16, 10812-10818, 2024.

[Aguirre2024] F. Aguirre, A. Sebastian, M. le Gallo, W. Song, T. Wang, J. Joshua Yang, W. Lu, M.-F. Chang, D. Ielmini, Y. Yang, A. Mehonic, A. Kenyon, M. Villena, J. Roldan, Y. Wu, H.-H. Hsu, N. Raghavan, J. Suñé, E. Miranda, A. Eltawil, G. Setti, K. Smagulova, K. Salama, O. Krestinskaya, X. Yan, K. Ang, S. Jain, S. Li, O. Alharbi, S. Pazos, M. Lanza, "Hardware implementation of memristor-based artificial neural networks", Nature communications, 15, 1974, 2014.

[Ambrogio2018] S. Ambrogio, et al., "Equivalent-accuracy accelerated neural-network training using analogue memory", Nature, 558, pp. 60–67, 2018.

[Brown1969] D. B. Brown, D.F. Shriver, "Structures and Solid-State Reactions of Prussian Blue Analogs Containing Chromium, Manganese, Iron, and Cobalt", Inorganic Chemistry, 8, 37–42, 1969.

[Bisquert2024] J. Bisquert, J. B. Roldán, E. Miranda, "Hysteresis in memristors produces a conduction inductance and a conduction capacitance effects", Physical Chemistry Chemical Physics, 26, 13804-13813, 2024.

[Buser1977] H.J. Buser, D. Schwarzenbach, W. Petter, A. Ludi, "The crystal structure of Prussian Blue: $Fe_4[Fe(CN)_6]_3 \cdot xH_2O$", Inorganic Chemistry, 16 (11), 2704-2710, 1977.

[Chiu2019] Y.-C. Chiu et al., "A 40 nm 2 Mb ReRAM macro with 85% reduction in forming time and 99% reduction in page-write time using auto-forming and auto-write schemes," in Proc. Symp. VLSI Technol., 232–233, 2019.

[Chou2020] C.-C. Chou et al., "A 22nm 96KX144 RRAM Macro with a Self-Tracking Reference and a Low Ripple Charge Pump to Achieve a Configurable Read Window and a Wide Operating Voltage Range," 2020 IEEE Symposium on VLSI Circuits, Honolulu, HI, USA, pp. 1-2, 2020.

[Chua1971] L.O. Chua, "Memristor – The Missing Circuit Element", IEEE Transactions on Circuit Theory, 18(5), pp. 507–519, 1971. https://doi.org/10.1109/TCT.1971.1083337

[Chua1976] L.O. Chua, Sung Mo Kang, "Memristive devices and systems", Proceedings of the IEEE, 64(2), pp. 209–223, 1976. https://doi.org/10.1109/PROC.1976.10092

[Chua2011] L.O. Chua, "Resistance switching memories are memristors", Applied Physics A, 102, 765–783, 2011.

[Danilin2015] S.N. Danilin, S.A. Shchanikov, "Neural network control over operation accuracy of memristor-based hardware", 2015 IEEE International conference on mechanical engineering, automation and control systems (MEACS), 1, 2015. DOI: 10.1109/MEACS.2015.7414916

[Faita2021] F.L. Faita, L.B. Avila, J.P.B. Silva, M.H. Boratto, C.C. Plá Cid, C.F.O. Graeff, M.J.M. Gomes, C.K. Müller, A.A. Pasa, "Abnormal resistive switching in electrodeposited Prussian White thin films", Journal of Alloys and Compounds, 896, 2021. 162971. doi.org/10.1016/j.jallcom.2021.162971.

[Grossi2016] A. Grossi et al., "Fundamental variability limits of filament-based RRAM," 2016 IEEE International Electron Devices Meeting (IEDM), San Francisco, CA, USA, pp. 4.7.1-4.7.4, 2016.

[Jain2019] P. Jain et al., "A 3.6 Mb 10.1 Mb/mm2 embedded non-volatile Re-RAM macro in 22nm FinFET technology with adaptive forming/set/reset schemes yielding down to 0.5 V with sensing time of 5ns at 0.7 V," 2019 IEEE International Solid-State Circuits Conference - (ISSCC), San Francisco, CA, USA, pp. 212-214,, 2019.

[Knot2021] A. C. Khot, T. D. Dongale, J. H. Park, A. V. Kesavan, T. G. Kim, "Experimental and Modeling Study of Metal–Insulator Interfaces to Control the Electronic Transport in Single Nanowire Memristive Devices", *ACS Applied Material Interfaces*, *13*, 5216, 2021.

[Knot2022] A. C. Khot, T. D. Dongale, K. A. Nirmal, J. H. Sung, H. J. Lee, R. D. Nikam, T. G. Kim, "Amorphous boron nitride memristive device for high-density memory and neuromorphic computing applications", *ACS Applied Material Interfaces*, *14*, 10546, 2022.

[Lanza2019] M. Lanza, H.-S. P. Wong, E. Pop, D. Ielmini, D. Strukov, B.C. Regan, L. Larcher, M.A. Villena, J.J. Yang, L. Goux, A. Belmonte, Y. Yang, F. M. Puglisi, J. Kang, B. Magyari-Köpe, E. Yalon, A. Kenyon, M. Buckwell, A. Mehonic, A. Shluger, H. Li, T.-H. Hou, B. Hudec, D. Akinwande, R. Ge, S. Ambrogio, J.B. Roldan, E. Miranda, J. Suñé, K.L. Pey, X. Wu, N. Raghavan, E. Wu, W.D. Lu, G. Navarro, W. Zhang, H. Wu, R. Li, A. Holleitner, U. Wurstbauer, M. Lemme, M. Liu, S. Long, Q. Liu, H. Lv, A. Padovani, P. Pavan, I. Valov, X. Jing, T. Han, K. Zhu, S. Chen, F. Hui, Y. Shi, "Recommended methods to study resistive switching devices", Advanced Electronics Materials, 5, 1800143, 2019.

[Lanza2022] Mario Lanza, Abu Sebastian, Wei D. Lu, Manuel Le Gallo, Meng-Fan Chang, Deji Akinwande, Francesco M. Puglisi, Husam N. Alshareef, Ming Liu, Juan B.

Roldan, "Memristive technologies for data storage, computation, encryption and radio-frequency communication", Science, 376, 6597, eabj9979, pp. 1-13, 2022.

[Lanza2025] M. Lanza, S. Pazos, F. Aguirre, A. Sebastian, M. Le Gallo, S. M. Alam, S. Ikegawa, J. J. Yang, E. Vianello, M.-F. Chang, G. Molas, I. Naveh, D. Ielmini, M. Liu, J. B. Roldán, "The growing memristor industry", Nature, 640, 613-622, 2025.

[Maldonado2022] D. Maldonado, S. Aldana, M.B. González, F. Jiménez-Molinos, M.J. Ibáñez, D. Barrera, F. Campabadal, J.B. Roldán, "Variability estimation in resistive switching devices, a numerical and kinetic Monte Carlo perspective ", Microelectronics Engineering, 257, p. 111736, 2022.

[Maldonado2022b] D. Maldonado, S. Aldana, M.B. González, F. Jiménez-Molinos, F. Campabadal, J.B. Roldán, "Parameter extraction techniques for the analysis and modeling of resistive memories", Microelectronics Engineering, 265, p. 111876, 2022.

[Maestro-Izquierdo2019] M. Maestro-Izquierdo, M.B. González, F. Campabadal, "Mimicking the spike-timing dependent plasticity in $HfO_2$-based memristors at multiple time scales", Microelectronics Engineering, 215, 111014, 2019.

[Mead1989] C. Mead and M. Ismail, "Analog VLSI Implementation of Neural Systems", Springer, 1989.

[Mikhaylov2020] A. Mikhaylov, A. Belov, D. Korolev, I. Antonov, V. Kotomina, A. Kotina, E. Gryaznov, A. Sharapov, M. Koryazhkina, R. Kryukov, S. Zubkov, A. Sushkov, D. Pavlov, S. Tikhov, O. Morozov, D. Tetelbaum, "Multilayer Metal-Oxide Memristive Device with Stabilized Resistive Switching" Advanced Materials Technologies, 5, 1900607, 2020.

[Milo2016] Milo, V., Pedretti, G., Carboni, R., Calderoni, A., Ramaswamy, N., Ambrogio, S., & Ielmini, D., "Demonstration of hybrid CMOS/RRAM neural networks with spike time/rate-dependent plasticity", 2016 IEEE International Electron Devices Meeting (IEDM), San Francisco, CA, USA, pp. 16.8.1-16.8.4, 2016.

[Mishchenko2022] M. Mishchenko, D. Bolshakov, V. Lukoyanov, D. Korolev, A.I. Belov, D. Guseinov, V. Matrosov, V. Kazantsev, A.N. Mikhaylov, "Inverted spike-rate-dependent plasticity due to charge traps in a metal-oxide memristive device" J. Phys. D: Appl. Phys., 2022.

[Nili2018] H. Nili et al., "Hardware-intrinsic security primitives enabled by analogue state and nonlinear conductance variations in integrated memristors," Nature Electronics, 1, 197–202, 2018.

[Nirmal2022] K. A. Nirmal, G. S. Nhivekar, A. C. Khot, T. D. Dongale, T. G. Kim, "Unraveling the Effect of the Water Content in the Electrolyte on the Resistive Switching Properties of Self-Assembled One-Dimensional Anodized $TiO_2$ Nanotubes", *Journal of Physical Chemistry Letters*, *13*, 7870, 2022.

[Pan2014] F. Pan, S. Gao, C. Chen, C. Song, F. Zeng, "Recent progress in resistive random access memories: materials, switching mechanisms and performance", Materials Science and Engineering R: Reports, 83, pp. 1-59, 2014.

[Perez2019] E. Pérez, D. Maldonado, C. Acal, J.E. Ruiz-Castro, F.J. Alonso, A.M. Aguilera, F. Jiménez-Molinos, Ch. Wenger, J.B. Roldán, "Analysis of the statistics of device-to-device and cycle-to-cycle variability in TiN/Ti/Al:$HfO_2$/TiN RRAMs", Microelectronics Engineering, 214, pp. 104-109, 2019.

[Perez-Bosch2021] E. Pérez-Bosch, R. Romero-Zaliz, E. Pérez, M. Kalishettyhalli, J. Reuben, M. A. Schubert, F. Jiménez-Molinos, J. B. Roldán, C. Wenger, "Toward reliable compact modeling of multilevel 1T-1R RRAM devices for neuromorphic systems", Electronics, 10, 645, 2021.

[Poblador2018] S. Poblador, M.B. González and F. Campabadal, "Investigation of the multilevel capability of TiN/Ti/HfO2/W resistive switching devices by sweep and pulse programming", Microelectronics Engineering, vol. 187-188, p. 148, 2018.

[Poblador2020] S. Poblador, M. Maestro-Izquierdo, M. Zabala, M. B. González, F. Campabadal "Methodology for the characterization and observation of filamentary spots in $HfO_x$-based memristor devices", Microelectronics Engineering, 223, p. 111232, 2020.

[Prezioso2015] M. Prezioso, F. Merrikh-Bayat, B. D. Hoskins, G. C. Adam, K. K. Likharev, D. B. Strukov "Training and operation of an integrated neuromorphic network based on metal-oxide memristors", Nature 521, pp. 61-64 (2015).

[Reuben2019] Reuben, J., Fey, D., Wenger, C., "A Modeling Methodology for Resistive RAM Based on Stanford-PKU Model With Extended Multilevel Capability", IEEE Transactions on Nanotechnology, 18, pp. 647-656, 2019.

[Roldan2021] J. B. Roldán, G. González-Cordero, R. Picos, E. Miranda, F. Palumbo, F. Jiménez-Molinos, E. Moreno, D. Maldonado, S. B. Baldomá, M. Moner Al Chawa, C. de Benito, S. G. Stavrinides, J. Suñé, L. O. Chua, "On the Thermal Models for Resistive Random Access Memory Circuit Simulation", Nanomaterials, 11, 1261, 2021.

[Roldan2022] J. B. Roldan, D. Maldonado, C. Aguilera-Pedregosa, E. Moreno, F. Aguirre, R. Romero-Zaliz, A. García-Vico, Y. Shen, Y. Yuan, M. Lanza, "Spiking neural networks based on two-dimensional materials", npj 2D Materials and Applications, 6, p. 63, 2022.

[Roldan2022b] Juan B Roldan, David Maldonado, C Aguilera-Pedregosa, Francisco J Alonso, Yiping Xiao, Yaqing Shen, Wenwen Zheng, Yue Yuan, Mario Lanza, "Modeling the variability of Au/Ti/h-BN/Au memristive devices", IEEE Transactions on Electron Devices, 70, 1533-1539, 2022.

[Roldan2023] J.B. Roldán, E. Miranda, D. Maldonado, A.N. Mikhaylov, N.V. Agudov, A.A. Dubkov, M. N. Koryazhkina, M.B. González, M.A. Villena, S. Poblador, M. Saludes-Tapia, R. Picos, F. Jiménez-Molinos, S. G. Stavrinides, E. Salvador, F.J. Alonso, F. Campabadal, B. Spagnolo, M. Lanza, L.O. Chua, "Variability in resistive memories", Advanced Intelligent Systems, p. 2200338, 2023.

[Romero-Zaliz2021] R. Romero-Zaliz, E. Pérez, F. Jiménez-Molinos, C. Wenger, J.B. Roldán, "Influence of variability on the performance of HfO2 memristor-based convolutional neural networks", Solid State Electronics, 185, 108064, 2021.

[Sangman2020] V. K. Sangwan, M. C. Hersam, "Neuromorphic nanoelectronic materials", *Nature Nanotechnology 2020 15:7*, *15*, 517, 2020.

[Sebastian2020] A. Sebastian, M. Le Gallo, R Khaddam-Aljameh et al., "Memory devices and applications for in-memory computing", Nature nanotechnology 15, pp. 529–544, 2020.

[Spiga2020] S. Spiga, A. Sebastian, D. Querlioz, B. Rajendran, "Memristive devices for brain-inspired computing", Elsevier, 2020.

[Tang2019] Tang, J.; Yuan, F.; Shen, X.; Wang, Z.; Rao, M.; He, Y.; Sun, Y.; Li, X.; Zhang, W.; Li, Y.; et al., "Bridging Biological and Artificial Neural Networks with Emerging Neuromorphic Devices: Fundamentals, Progress, and Challenges", Advanced Materials, 31, p. 1902761, 2019.

[Villena2014] M. A. Villena, J. B. Roldán, F. Jimenez-Molinos, J. Suñé, S. Long, E. Miranda, M. Liu, "A comprehensive analysis on progressive reset transitions in RRAMs", Journal of Physics D: applied physics, 47(20), 205102, 2014.

[Wen2021] C. Wen, X. Li, T. Zanotti, F. M. Puglisi, Y. Shi, F. Saiz, A. Antidormi, S. Roche, W. Zheng, X. Liang, J. Hu, S. Duhm, J. B. Roldan, T. Wu, V. Chen, E. Pop, B. Garrido, K. Zhu, F. Hui, M. Lanza, "Advanced data encryption using two-dimensional materials", Advanced Materials, 2100185, 1-12, 2021.

[Wong2012] H. S. P. Wong, H. Y. Lee, S. Yu. Y.-S. Chen, Y. Wu, P.-S. Chen, B. Lee, F. T. Chen and M.-J. Tsai. "*Metal-Oxide RRAM*" Proceedings of the IEEE, Vol. 100, Nº. 6, June, 2012.

[Yu2021] S. Yu, H. Jiang, S. Huang, X. Peng, A. Lu, "Computing-in-memory chips for deep learning: recent trends and prospects", IEEE circuits and systems magazine, pp. 31–56, 2021.

[Zhu2023] K. Zhu, S. Pazos, F. Aguirre, Y. Shen, Y. Yuan, W. Zheng, O. Alharbi, M. A. Villena, B. Fang, X. Li, A. Milozzi, M. Farronato, M. Muñoz-Rojo, T, Wang, R. Li, H. Fariborzi, J. B. Roldan, G. Benstetter, X. Zhang, H. Alshareef, T. Grasser, H. Wu, D. Ielmini, M. Lanza, "Hybrid 2D/CMOS microchips for memristive applications", Nature, 618, 57–62, 2023.